# Application of *R*-matrix and Lagrange-mesh methods to nuclear transfer reactions

Shubhchintak[1*] and P. Descouvemont[2]

Physique Nucléaire Théorique et Physique Mathématique, C. P. 229, Université Libre de Bruxelles (ULB), B 1050 Brussels, Belgium
[1*]Email: shubhchintak@ulb.be (corresponding author)
[2]Email: pierre.descouvemont@ulb.be

ABSTRACT

**Background:** Nuclear transfer reactions are a useful tool to study the structure of a nucleus. For reactions involving weekly bound nuclei, breakup effects can play significant role and theoretical calculations can be computational expensive in such cases.
**Purpose:** To utilize the Lagrange-mesh and R-matrix methods for nuclear transfer reactions.
**Methods:** We use the adiabatic distorted wave approximation (ADWA) method which can approximately treats the breakup effects in a simpler manner. In our approach, we apply the *R*-matrix method combining it with the Lagrange-mesh method, which is known to provide the fast and accurate computations.
**Results:** As a test case, we calculate the angular distribution of the cross sections for the $^{54}$Fe(d, p)$^{55}$Fe reaction, where deuteron breakup effects play important role.
**Conclusions:** We show that these methods work well in the ADWA framework, and we look forward to applying these methods in coupled channel calculations.

## 1. Introduction

Nuclear transfer reactions involving the transfer of particles (or a cluster) among the projectile and target, provide a useful tool to investigate the structure of a nucleus [1-2]. This is possible because of the dependence of transfer cross sections on the structure of projectile and residual nucleus. Comparison of the measured cross sections with the calculated ones, can provides information like angular momentum ($\ell$), spin-parity ($J^\pi$), spectroscopic factor (SF) or the asymptotic normalisation coefficient (ANC) of the populated state of a residual nucleus. In case, the final state is a resonance then the information about the width of resonance can be obtained by knowing the SF or the ANC of that particular state. These information like, spin-parity, energy, widths etc. are often required in several estimations of astrophysical reaction rates and in those cases transfer reactions are used as an indirect tool for the study of such reactions. For example, reactions involving the transfer of α cluster such as ($^7$Li, t) and ($^6$Li, d) have been used to study the (α, γ) and (α, n) reactions. See for example, Ref. [3] for more details.

There are different theoretical models for transfer reactions involving formalisms like, distorted wave approximation (DWBA), adiabatic method, continuum discretised coupled channel (CDCC) method and Faddeev method. Among all these, DWBA is the simplest approximation where the transfer is consider as one-step process and since a long time it is being used for the analysis of experimental data. Modern calculations, like in the CDCC and Faddeev's methods are more demanding in terms of the computer capabilities and hence require efficient numerical techniques. In this regard, as a first step, recently we have applied the *R*-matrix method [4] along with the Lagrange-mesh method [5] in the DWBA framework of transfer reactions [6–7]. We have shown that these methods lead to the faster and accurate numerical computations. For weekly bound nuclei it becomes important to take the breakup channels into account. In such cases, DWBA which considers only the elastic channel, may not works well in explaining the data. Latter three methods mentioned above, on the other hand, can treat breakup effects effectively. The CDCC and Faddeev's method are more advanced methods but they are computationally expensive, whereas the adiabatic method which includes breakup channels in an approximate way, is known to provide similar results to those from CDCC method at higher energies but at lower energies they may differ.

In the adiabatic approximation, it is assumed that at relatively high incident energy the projectile is frozen during the collision. This is justified if the binding energy of the projectile is much smaller than the scattering energy. This method has been used in several transfer reaction studies involving deuteron (see for example Refs.[8-10]) and was initially proposed by Johnson and Soper [11], where they used zero-range form of the deuteron binding potential. A finite-range version of this method was introduced by Johnson and Tandy (JT) in Ref. [12], where the scattering matrix of the transfer process, finally, transformed to a simpler form like the one in the DWBA. For more details, one can see for example Refs. [12-13]. Due to its simplicity we use JT-form of the adiabatic method in the present study and discuss a case of (d, p) reaction. Our aim is to utilize the Lagrange-mesh and $R$-matrix method as we did in Refs. [6-7] in the DWBA framework of transfer reactions.

This paper is organized as follow. In section 2, we briefly describe the adiabatic method for transfer reactions and also discuss in brief the $R$-matrix and the Lagrange mesh methods. Section 3, consists of our results and discussions where we present the transfer cross section for the $^{54}$Fe(d, p)$^{55}$Fe and then finally we conclude in Section 4.

## 2. Formalism

### 2.1 Outline of the ADWA

We consider a transfer reaction $d(p+n) \to B(t+n) + p$, where neutron ($n$) is transferred from the incident deuteron ($d$) to the target $t$ and form a residual nucleus $B$ in the final state along with the outgoing proton ($p$). Fig. 1, shows various coordinates involved in the process.

Following Refs. [6,13], the scattering matrix for the above stripping reaction can be written as

$$U_{\alpha\beta}^{J\pi} = -\frac{i}{\hbar}\left\langle\chi_{L_B}^{J\pi}(\boldsymbol{R'})\phi_{\ell_B}^{I_B}(\boldsymbol{r}_B)\big|V_{np}\big|\Psi_\alpha^{(+)}(\boldsymbol{R},\boldsymbol{r}_d)\right\rangle, \tag{1}$$

where $\phi_{\ell_B}$ and $\chi_{L_B}$ are the bound and scattering state wave functions of nucleus $B$ with $\ell_B$ and $L_B$ as their respective angular momenta. $I_B$ represents the spin of nucleus $B$ and $J$ is the total angular momentum with parity $\pi$. $V_{np}$ is the neutron-proton interaction and $\Psi_\alpha^{(+)}(R, r_d)$ is the exact three-body wave function in the incident channel. Labels α and β stand for $(L_d, \ell_d, I_d)$ and $(L_B, \ell_B, I_B)$, respectively.

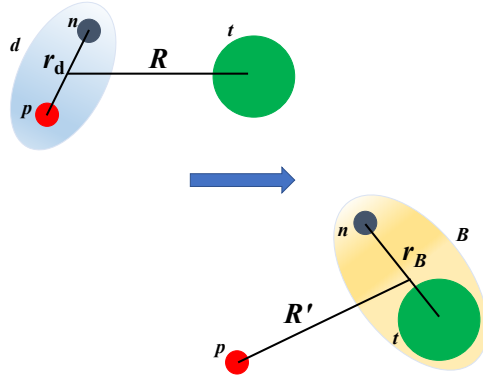

**Figure 1**: Schematic diagram of the deuteron stripping process where a neutron $n$ is transferred to the target $t$ and a nucleus $B$ is formed along with the outgoing proton $p$.

The three-body wave function $\Psi_\alpha^{(+)}(\boldsymbol{R},\boldsymbol{r}_d)$ is a solution of the inhomogeneous differential equation (see Ref. [12])

$$\left[E + i\epsilon - T_{r_d} - T_R - V_{np}(\boldsymbol{r}_d) - U_{nt}\left(\boldsymbol{R} - \frac{1}{2}\boldsymbol{r}_d\right) - U_{pt}\left(\boldsymbol{R} + \frac{1}{2}\boldsymbol{r}_d\right)\right]\Psi^{(+)}(\boldsymbol{R},\boldsymbol{r}_d) = i\epsilon\phi_d(\boldsymbol{r}_d)e^{i\boldsymbol{K}_d\cdot\boldsymbol{R}}, \tag{2}$$

where $E = E_d - \epsilon_d$, with $E$ and $\epsilon_d$, respectively, as the kinetic and binding energies of the deuteron. $T_{r_d}$ and $T_R$ are the kinetic energy operators for the relative motion of neutron-proton and of deuteron-target, respectively. $U_{nt}$ and $U_{pt}$ are the neutron-target and proton-target interactions at half the incident energy, respectively and $K_d$ is the deuteron wave number. The term on the r.h.s. ensures the incoming boundary condition in the deuteron channel, where $\phi_d$ is the internal wave function of the deuteron.

In the JT adiabatic method, main emphasis is given to the product $V_{np}|\Psi^{(+)}(\mathbf{R}, \mathbf{r}_d)\rangle$ required in Eq. (1) and for this the three-body wave function $\Psi^{(+)}(\mathbf{R}, \mathbf{r}_d)$ is expanded in terms of Weinberg basis $\phi_i(\mathbf{r}_d)$ which are given by

$$[-\epsilon_d - T_r - \alpha_i V_{np}(\mathbf{r}_d)]\phi_i(\mathbf{r}_d) = 0, \tag{3}$$

where, $\alpha_i$ with $i = 1, 2, 3 \ldots$, are the eigen values of Eq. (3). These basis functions form a complete set of square integrable functions and are orthogonal in the sense that $\langle\phi_i|V_{np}|\phi_j\rangle = -\delta_{ij}$. States $\phi_i$ become increasingly oscillatory with increase in $i$ (as $\alpha_i$ increase monotonically) and possess $i$ nodes within the range of $V_{np}$, whereas at large $r$, they decay exponentially. Apart from normalization, $\phi_1$ is same as the deuteron ground state wave function, where $\alpha_1 = 1$.

It has been shown in Refs. [12] that expanding the three-body wave function in terms of Weinberg basis as

$$\Psi^{(+)}(\mathbf{R}, \mathbf{r}_d) = \sum_{i=1}^{\infty} \phi_i(\mathbf{r}_d)\chi_i(\mathbf{R}), \tag{4}$$

with $\chi_i(\mathbf{R}) = -\langle\phi_i|V_{np}|\Psi^{(+)}\rangle$, Eq. (2) will change to a set of coupled differential equation in $\chi_i(\mathbf{R})$ which can be solved exactly [14]. Each channel wave function in that case then contains the breakup effects, however, the energy of the first channel ($i = 1$) will remain unchanged at the elastic deuteron value $E_d$ [12].

If one keeps only the first term with $\alpha_1 = 1$, which is the JT-ADWA, an optical model like equation in $\chi_i(\mathbf{R})$ is obtained and is given by

$$[E_d + i\epsilon - T_R - U_{11}(\mathbf{R})]\chi_i^{JT}(\mathbf{R}) = i\epsilon N_d e^{i\mathbf{K}_d \cdot \mathbf{R}}, \tag{5}$$

where the superscript JT stands for Johnson-Tandy ADWA and the normalization coefficient $N_d = -\langle\phi_1|V_{np}|\phi_d\rangle$. The potential $U_{11}$ is given in terms of sum of neutron and proton potentials and can be written as

$$U_{11}(\mathbf{R}) = U_{ADWA}(\mathbf{R}) = \frac{\langle\phi_d(\mathbf{r}_d)|V_{np}(U_{nt} + U_{pt})|\phi_d(\mathbf{r}_d)\rangle}{\langle\phi_d(\mathbf{r}_d)|V_{np}|\phi_d(\mathbf{r}_d)\rangle}. \tag{6}$$

The zero-range limit of the above potential gives the Johnson-Soper potential [11]. The scattering matrix [Eq. (1)], then simplifies as

$$U_{\alpha\beta}^{J\pi} = -\frac{i}{\hbar}\langle\chi_{L_B}^{J\pi}(\mathbf{R}')\phi_{\ell_B}^{I_B}(\mathbf{r}_B)|V_{np}|\phi_d(\mathbf{r}_d)[\chi_1^{J\pi}(\mathbf{R})/N_d]\rangle, \tag{7}$$

which is quite similar to the scattering matrix in the DWBA framework (see for example Ref. [6-7]), except that the distorted wave $[\chi_1^{JT}(\mathbf{R})/N_d]$ is now generated by the potential $U_{11}$ [Eq. (6)] instead of the deuteron optical potential. We use the $R$-matrix method to calculate the distorted wave functions $\chi_i$ in Eq. (7) which are obtained in many other works mainly by the finite-difference method.

### 2.2 R-matrix and the Lagrange-mesh methods

Here we discuss these methods in brief, but for more details one is referred to Refs. [4-6]. In the $R$-matrix method, the configuration space is divided into two parts, the internal region (with radius $a$) and the external region. The channel radius $a$ should be selected as large enough so that the nuclear interactions are negligible. The wave function in the internal region ($R \leq a$) is expanded over a set of $N$ basis functions as

$$\chi_{int}^L(R) = \sum_{j=1}^{N} c_j^L \varphi_j(R), \tag{8}$$

where $c_j^L$ are the expansion coefficients and $\varphi_j(R)$ are the basis function. In our approach we use the Lagrange functions as the basis, the importance of this choice will be discussed in coming paragraphs. For simplicity we omit other indices and only keep the angular momentum $L$. In the external region ($R > a$), the wave function takes the form

$$\chi_{ext}^L(R) = \frac{1}{\sqrt{v}}[I_L(kR) - U_L O_L(kR)], \tag{9}$$

where $I_L(kR)$ and $O_L(kR)$ are the incoming and outgoing Coulomb functions. $U_L$ is the scattering matrix for the elastic scattering and $v$ and $k$ are the velocity and wave number of the concerned particle. Note that the matrix elements of the kinetic energy are not Hermitian as the basis functions are valid only in the region $[0, a]$. To solve this problem one use the Bloch operator [15], which apart from ensuring the Hermiticity of the Hamiltonian, also leads to the continuity of the derivative of the wave function i.e. $\chi_{int}^L{'}(a) = \chi_{ext}^L{'}(a)$.

Following Refs. [4,6], one can then calculate the scattering matrix $U_L$ [in Eq. (9)] from the continuity condition, which is then used to calculate the wave function. The principle of the method is that from the properties of the Hamiltonian in the internal region the $R$-matrix can be calculated (defined as the reciprocal of the logarithmic derivative of the wave function at the channel radius) which is then used to determine the scattering matrix in the external region.

Now we come to the basis functions $\varphi_j(R)$, which as mentioned above we choose as Lagrange functions. These are $N$ infinitely differentiable functions which form an orthonormal set. They have the special property that they vanish at all mesh points except one and have a Gauss quadrature associated with the mesh. Use of these functions transformed the Schrödinger equation into a matrix form which can be easily solved. For more details one is referred to Ref. [5]. With these functions, matrix elements of the potential take a diagonal form which simplifies the calculations a lot. Similarly, the matrix elements of the kinetic energy are also simplified and can be easily calculated [4,5].

With the above mentioned procedure and following Ref. [6], we then simplify the scattering matrix [Eq. (7)] which is further computed to obtain the cross sections. As discussed in Refs. [6-7], typically $N \approx 30 - 40$ basis functions are sufficient to achieve the convergence which are significantly lesser than the number of points needed in other methods, like the finite difference method, where normally around 500 points are used. This makes the method very efficient and speedup our computations. It is worth mentioning that the channel radius here is not a parameter and the scattering matrix should not depends on it. Any value of $R$, so that the nuclear interaction becomes negligible, can be used. However, a large channel radius needs large number of basis functions $\phi_i(R)$ and hence the computer times. Therefore, as a compromise one needs to choose the channel radius as small as possible.

## 3. Results and discussions

As a test case we calculate the cross section for the transfer reaction $^{54}$Fe(d, p)$^{55}$Fe(1/2$^-$) at $E_d = 23$ MeV, where a neutron transferred to $^{54}$Fe leads to the formation of $^{55}$Fe in its first excited state (1/2$^-$, $E_x = 0.411$ MeV). Before presenting the results of our calculations it is important to mention various parameters used in the calculations. These include $R$-matrix parameters and various optical potentials. We use a channel radius $a = 20$ fm, and number of basis N = 80. These are chosen large enough to ensure the convergence. In Refs. [6,7], a procedure to check the convergence of transfer cross sections with respect to these parameters has been presented and same is followed here. We use integer masses and the constant $\hbar^2/2M_N = 20.9$ MeV.fm² is used ($M_N$ is the nucleon mass).

Standard Gaussian potential is used to calculate the deuteron ground-state wave function ($s$ state), with the following parameters

$$V_{np}(r) = -72.66 \exp[-(r/1.484)^2]. \tag{10}$$

Bound state of $^{55}$Fe(1/2$^-$) is obtained by using the Woods-Saxon potential with parameters $r_0 = 1.25$ fm, $a = 0.65$ fm and $V_0 = 52.5$ MeV.

Optical potentials of proton and neutron are obtained from global parametrization of Ref. [16], which are of the form

$$U(r) = -V_r f(r, R_r, a_r) + V_c(r) - iW_v f(r, R_v, a_v) - i W_s g(r, R_s, a_s), \tag{11}$$

where $V_c$ is the Coulomb potential of uniformly charged sphere with radius $R_c$. $f(r, R, a) = 1/[1 + \exp\left(\frac{r-R}{a}\right)]$ and the imaginary part contains a volume term and a surface term with

$$g(r, R_s, a_s) = -4\, a_s \frac{d}{dr} f(r, R_s, a_s). \tag{12}$$

In Fig. 2, we plot the angular distribution of the cross section as a function of angle in the centre of mass frame (solid line). We compare the calculations with the experimental data of Ref. [17]. This also show that our approach of using Lagrange-mesh and R-matrix methods works well. For a comparison we also plot the cross sections calculated using DWBA (dashed line) of Ref. [6]. In this case the deuteron scattering wave function is calculated by using the optical potential given in Ref. [18], having the parameters $V_r = 105$ MeV, $R_r = 3.85\, fm$, $a_r = 0.86\, fm$, $W_s = 15$ MeV, $R_s = 5.37\, fm$, $a_s = 0.65\, fm$ and $R_c$ is taken as 4.9 fm, whereas global potentials of Ref. [16] are used for the proton.

It is clear from the figure that both calculations give similar results up to the first minima and then they start differing. So, the extracted SFs (which are normally obtained by fitting the data up to the first minima) in this case may not change much when using either of these frameworks. At large angles where deuteron breakup effects play their role, DWBA cross sections deviate from the data whereas, the JT-ADWA calculations which include breakup effects, can nicely generate all the oscillations of the experimental cross sections and has a reasonable agreement with the data. We also tested some other optical potentials of the neutron and proton, and they also give similar results with slight change in the cross sections.

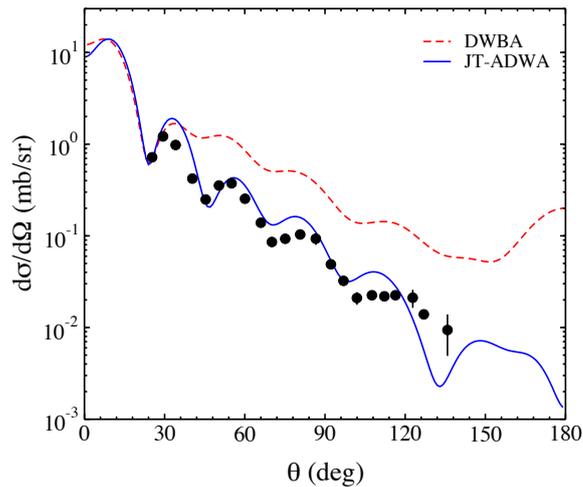

**Figure 2**: Angular distribution of the cross sections for the $^{54}$Fe(d, p)$^{55}$Fe(1/2-) at $E_d = 23$ MeV energy. Dashed and solid lines are calculations with DWBA and adiabatic approximation, respectively. Experimental data are taken from Ref. [17].

## 4. Conclusion

We have applied the R-matrix and the Lagrange mesh methods to the transfer reactions involving deuteron in the Johnson-Tandy ADWA framework. This method includes the deuteron breakup effects in an approximate way and at relatively high energies can gives cross sections in agreement with the other more advanced frameworks of transfer reactions, such as the CDCC. As a test case, we considered the $^{54}$Fe(d, p)$^{55}$Fe(1/2-) reaction at 23 MeV, where first excited state of $^{55}$Fe is formed. We have also compared the ADWA calculations with the ones from the DWBA framework. Their comparison shows the importance of deuteron breakup channels in this case. Calculations with the ADWA give a nice agreement with the data. This also shows that the method works well, and the use of Lagrange-mesh and R-matrix methods brings additional simplifications. One needs only few points to calculate wave functions and the radial and angular integrals. This encourages us to utilize these methods for the computationally expensive frameworks of transfer reactions such as the CDCC method.

### Acknowledgements

This work has received funding from the European Union's Horizon 2020 research and innovation program under the Marie Skłodowska-Curie grant agreement No 801505.